\documentstyle[12pt]{article}
\def\mbf#1{\hbox{\boldmath $#1$}}

\def\citen#1{\cite{#1}}

\def\bq{{\mbf q}}

\def\br{{\mbf r}}

\def\bx{{\mbf x}}

\def\bX{{\mbf X}}

\def\CP{{\cal P}}

\def\CA{{\cal A}}

\def\SM3{\Sigma N (3/2)}
\def\SN1{\Sigma N (1/2)}

\def\TS1{\hbox{}^3S_1}
\def\TD1{\hbox{}^3D_1}
\def\CA{{\cal A}}

\def\CP{{\cal P}}

\def\CA{{\cal A}}

\def\aagp{{\alpha \alpha^\prime}}

\def\eq#1{Eq.~(\ref{#1})}

\def\ph{\phi^{\rm int}}
\def\EKi{E^{\rm K}_{\rm int}}
\def\EVi{E^{\rm V}_{\rm int}}
\title{$G$-Matrix Equation in the Resonating-Group Method}

\author{
Yoshikazu {\sc Fujiwara}\thanks{e-mail:
fujiwara@ruby.scphys.kyoto-u.ac.jp}, Department of Physics,
Kyoto University \\
Kyoto 606-8502, Japan
\and
Michio {\sc Kohno}, Physics Division, Kyushu Dental College \\
Kitakyushu 803-8580, Japan
\and
Choki {\sc Nakamoto}, Suzuka National College of Technology \\
Suzuka 510-0294, Japan
\and
Yasuyuki {\sc Suzuki}, Department of Physics, Niigata University \\
Niigata 950-2181, Japan
}

\date{28 June 2000}

\begin{document}

\maketitle

\vspace{-2mm}

\begin{center}
{\bf abstract}
\end{center}
The $G$-matrix equation is most straightforwardly formulated
in the resonating-group method if the quark-exchange kernel is
directly used as the driving term for the infinite sum
of all the ladder diagrams.
The inherent energy-dependence involved in the exchange term
of the normalization kernel plays the essential role to define
the off-shell $T$-matrix uniquely
when the complete Pauli-forbidden state exists.
We analyze this using a simple solvable model
with no quark-quark interaction, and calculating
the most general $T$-matrix
in the formulation developed by Noyes and Kowalski.
This formulation gives a certain condition
for the existence of the solution
in the Lippmann-Schwinger resonating-group method.
A new procedure to deal with the corrections for the reduced masses
and the internal-energy terms
in the $\Lambda N$-$\Sigma N$ coupled-channel
resonating-group equation is proposed.

\normalsize

\section{Introduction}

The $G$-matrix formalism is one of the best known frameworks
to study the effective baryon-baryon interaction
in a nuclear medium. \cite{LOBT,DAY}
When the free baryon-baryon scattering
is described by the non-relativistic Schr{\" o}dinger equation
with a simple local potential, the derivation of the $G$-matrix
equation is straightforward even for the most general
off-shell $G$-matrix used in the many-body calculations.
This is not the case if the basic baryon-baryon interaction is
formulated as a composite-particle interaction in the framework
of the resonating-group method (RGM). \cite{WH37,WT77}
It is well known that the relative wave-function
between clusters has a different normalization property
from the wave-function of the ordinary Schr{\" o}dinger equation,
owing to the antisymmetrization
of constituent quarks ($q$). \cite{SA77}
It is, therefore, claimed that the quark-model potential
derived from the $(3q)$-$(3q)$ RGM \cite{OY84} should be defined by
rewriting the original RGM equation
to the Schr{\" o}dinger-type equation.
One of the merits of this method is that
it is possible to eliminate the explicit energy dependence
of the quark-exchange kernel,
which inherently appears in the RGM equation.
The essential point of this procedure is
to renormalize the RGM relative wave-function $\chi (\br)$ by
the exchange normalization
kernel $K$ as $\psi (\br)=\sqrt{1-K}\chi (\br)$.
This decent prescription, however, needs a special care when a complete
Pauli-forbidden state exists.

In this paper, we discuss roles of the Pauli-forbidden state
in the formulation of the $G$-matrix equation for the RGM. 
It is shown that the direct use of the quark-exchange kernel
for the driving term of the $G$-matrix or $T$-matrix equation
is the simplest and the most natural procedure
in the sense that the orthogonality condition \cite{SA68}
to the complete Pauli-forbidden state is automatically incorporated
in the structure of the exchange kernel.
The concept of the orthogonality need not be applied only
to the relative wave-function, but could also be applied
to any other physical quantities appearing
in the $T$-matrix formulation of the scattering
in a nuclear medium. The energy dependence involved
in the normalization exchange kernel is not
an unfavorable feature, but is essential to represent the effect
of the compositeness of the nucleon clusters.
To show these, we use a simplest version
of the Saito's orthogonality condition
model (OCM) \cite{SA68,SA77} with no $q$-$q$ interaction.
In this case, the exact solution
of the complete off-shell $T$-matrix is analytically given.
If we solve this problem
in the Noyes \cite{NO65} and Kowalski \cite{KO65} method,
it is easy to see what kind of condition is necessary
to guarantee the existence of the solution of the basic
Lippmann-Schwinger-type equation.
In the general RGM equation, this condition is automatically satisfied.
When we need to modify the exchange kernel,
the modification should be made in such a way that
this condition is still satisfied.
An example of this kind of modification is
the correction of the reduced masses
and a small readjustment of the threshold
energies in the coupled-channel RGM (CCRGM). 
One can preserve a realistic kinematics
of the baryon-baryon scatterings
even in a very rigorous framework of the CCRGM.  

In the next section, we first briefly illustrate the RGM formalism
to show the notation used in this paper.
An orthogonality condition model is introduced
as a simplified version of the RGM.
The analytic solution of the $T$-matrix equation
is given in a heuristic way.
Next we consider the basic equations for the most general $T$-matrix
in the Noyes and Kowalski method.
The orthogonality relations to the Pauli-forbidden
state are analyzed in $\S$3 with respect to the
simplest version of the OCM.
The modification of the exchange kernel
of the RGM equation is  discussed in $\S$4. 
The final section is devoted to discussion and a brief summary.

\bigskip


\section{Formulation}

\subsection{RGM equation}

The RGM equation for the relative-motion wave-function $\chi (\br)$ is
usually formulated from the variational equation \cite{WT77}
\begin{equation}
\langle \ph \vert E-H \vert \CA \{ \ph \chi \} \rangle =0\ ,
\label{rgm1}
\end{equation}
where $\ph$ is an appropriate internal cluster function
and the total Hamiltonian consists of 
\begin{equation}
H=\sum^6_{i=1} t_i-T_G+\sum^6_{i<j} v_{ij}\ .
\label{rgm2}
\end{equation}
Here we particularly consider $(3q)$-$(3q)$ RGM \cite{OY84} for
the baryon-baryon interaction.
The normalization kernel $N$ stands for
\begin{equation}
N=\langle \ph \vert \CA \vert \ph \rangle = 1-K\ , 
\label{rgm3}
\end{equation}
where the exchange normalization kernel $K$ sometimes
allows a complete Pauli-forbidden state $\vert u \rangle$ satisfying
\begin{equation}
K \vert u \rangle=\vert u \rangle \qquad \hbox{and} \qquad
\CA \{ \ph u \}=0\ .
\label{rgm4}
\end{equation}
We use a common notation, $\Lambda=1-\vert u \rangle
\langle u \vert$, to denote
the projection operator on the Pauli-allowed space. \cite{SA77}
For the direct term, we express $H$ as
\begin{equation}
H=H_{\rm int}+H_0+\sum^3_{i=1} \sum^6_{j=4} v_{ij}\ ,
\label{rgm5}
\end{equation}
and define the internal-energy term $E_{\rm int}$ and
the direct potential $V_{\rm D}$ by
\begin{eqnarray}
& & \langle \ph \vert H_{\rm int} \vert \ph \rangle = E_{\rm int}=\EKi
+\EVi \ ,\nonumber \\
& & \langle \ph \vert \sum^3_{i=1} \sum^6_{j=4} v_{ij}
\vert \ph \rangle = V_{\rm D} \ .
\label{rgm6}
\end{eqnarray}
For the exchange term, we use the notation
\begin{eqnarray}
G & = & G^{\rm K}+G^{\rm V}\ ,\nonumber \\ 
G^{\rm K} & = & \langle \ph \vert ( \sum^6_{i=1} t_i-T_G )
(\CA-1) \vert \ph \rangle
+\EKi K\ ,\nonumber \\
G^{\rm V} & = & \langle \ph \vert ( \sum^6_{i<j} v_{ij} )
(\CA-1) \vert \ph \rangle 
+\EVi K\ ,
\label{rgm7}
\end{eqnarray}
The exchange kinetic-energy kernel $G^{\rm K}$ is symmetric,
since $\EKi$ is common for all channels,\footnote{Here
we use the simplest center-of-mass 
coordinate $\bX=(\bx_1+\bx_2+\bx_3)/3$ for the $(3q)$ clusters and
assume that the effect of the flavor symmetry
breaking is respected only in the original Hamiltonian $H$.}
while $G^{\rm V}+\EVi K$ is symmetric for
the exchange interaction kernel.
Using these notations, \eq{rgm1} is converted
into the usual RGM equation:
\begin{eqnarray}
& & (\varepsilon - H_0-V_{\rm RGM}) \chi =0 \ ,
\label{rgm8}
\end{eqnarray}
with $\varepsilon=E-E_{\rm int}$ and $V_{\rm RGM}=V_{\rm D}
+G+\varepsilon K$.

The essential feature of the RGM equation, \eq{rgm8},
is the existence of the trivial
solution $\chi=u$; namely, we can write \eq{rgm8} as
\begin{equation}
\Lambda (\varepsilon - H_0-V_{\rm RGM} ) \Lambda \chi = 0\ .
\label{rgm9}
\end{equation}
This equation is rewritten as 
\begin{equation}
(\varepsilon - H_0-V) \chi = 0 \ ,
\label{rgm10}
\end{equation}
with
\begin{eqnarray}
& & V=V(\varepsilon)+v \ ,\nonumber \\
& & V(\varepsilon)=(\varepsilon-H_0)-\Lambda (\varepsilon-H_0)
\Lambda\ ,\nonumber \\
& & v=\Lambda V_{\rm RGM}\Lambda =\Lambda (V_{\rm D}
+G+\varepsilon K)\Lambda\ .
\label{rgm11}
\end{eqnarray}
We note simple relations,
$\Lambda V(\varepsilon)\Lambda=0$ and $v=\Lambda v \Lambda$.
Actually, $v$ also has a weak $\varepsilon$ dependence.
If the effect of the Pauli principle has a simple structure
such as in the deuteron-deuteron system,
the $\Lambda K \Lambda$ term exactly vanishes.
In this particular case, we can also show
that $G^{\rm K} \sim \Lambda H_0 \Lambda-H_0$  and $\Lambda
G^{\rm K} \Lambda \sim 0$.
This implies that $V(\varepsilon)$ term in \eq{rgm11} stands for
the dominant part of $G^{\rm K}+\varepsilon K$ in the
original RGM kernel $V_{\rm RGM}$,
and $v \sim \Lambda (V_{\rm D}+G^{\rm V})\Lambda$.

From these observations, we can conclude that the most essential part
of the Pauli principle of the RGM equation is already retained
in a simple OCM-type equation
\begin{equation}
(\varepsilon - H_0) \psi = V(\varepsilon) \psi \ ,
\label{rgm12}
\end{equation}
and the general RGM equation, \eq{rgm8}, can be obtained by adding
to $V(\varepsilon)$ the potential term $v$ having
the property $v=\Lambda v \Lambda$.
We will consider the $T$-matrix of \eq{rgm12} in the next
subsection.

\bigskip

\subsection{$T$-matrix of the simple OCM equation}

In this subsection we derive a complete off-shell $T$-matrix
for the simple OCM equation \cite{SA68,SA77}
\begin{equation}
\Lambda (\varepsilon - H_0) \Lambda \psi = 0 \ ,
\label{ocm1}
\end{equation}
which is equivalent to \eq{rgm12}.\footnote{The on-shell $T$-matrix
for more general OCM equations than \protect\eq{ocm1} with
local potentials has been extensively studied by many authors.
See, for example, Refs.\,\citen{OK72}, \citen{EN74},
\citen{GL76} and \citen{SU78}.}
Since the solution of \eq{ocm1} has an ambiguity of $u$,
we first consider a more general equation
\begin{equation}
(\omega - H_0) \psi = V(\varepsilon) \psi
\qquad \hbox{with} \qquad \omega \neq \varepsilon \ ,
\label{ocm2}
\end{equation}
and take the limit $\omega \rightarrow \varepsilon$ in the
final expression.
One can formulate this in two ways.
The first method is to consider the solution of
\begin{equation}
(\omega - H_0) \psi = V(\varepsilon) \psi \qquad \hbox{with} \qquad
(\omega - H_0) |\omega \rangle = 0 \ ,
\label{ocm3}
\end{equation}
and the second method is to use
\begin{equation}
(\omega - H_0) (\psi-\phi) = V(\varepsilon) \psi \qquad
\hbox{with} \qquad (\varepsilon - H_0) |\phi \rangle = 0 \ .
\label{ocm4}
\end{equation}
The latter equation is motivated by the correlation function
technique for the $G$-matrix,
for which $\chi=\phi-\psi$ corresponds to the so-called
defect function. In this case, the starting
energy $\omega$ is usually negative.
For $\omega \neq \varepsilon$ one can easily prove $\langle u|\psi
\rangle=0$ for the solution of \eq{ocm3} and $\langle u|\chi \rangle=0$
for the solution of \eq{ocm4}.

Both equations, Eqs.\,(\ref{ocm3}) and (\ref{ocm4}), lead to the same
definition for the $T$-matrix, given by
\begin{eqnarray}
& & T(\omega,\varepsilon)=V(\varepsilon)+V(\varepsilon) G^{(+)}_0(\omega)
T(\omega,\varepsilon) \nonumber \\
& & \hbox{with} \qquad G^{(+)}_0(\omega)={1 \over \omega-H_0+i0} \ .
\label{ocm5}
\end{eqnarray}
The solution of \eq{ocm5} is derived as follows.
We first define the free Green function in the allowed space by 
\begin{eqnarray}
G_{\Lambda}(\omega)=G^{(+)}_0(\omega)-G^{(+)}_0(\omega)|u\rangle
{1 \over \langle u|G^{(+)}_0(\omega)|u \rangle}
\langle u|G^{(+)}_0(\omega)\ ,
\label{ocm6}
\end{eqnarray}
which satisfies
\begin{eqnarray}
\Lambda (\omega-H_0) \Lambda G_{\Lambda}(\omega)
=G_{\Lambda}(\omega) \Lambda (\omega-H_0) \Lambda = \Lambda  \ .
\label{ocm7}
\end{eqnarray}
The Green function for $\omega-H_0-V(\varepsilon)$ can
be easily found to be
\begin{eqnarray}
G(\omega,\varepsilon)={1 \over \omega-H_0-V(\varepsilon)}
=G_{\Lambda}(\omega)+|u\rangle {1 \over \omega-\varepsilon}\langle u|\ .
\label{ocm8}
\end{eqnarray}
Since this $G(\omega,\varepsilon)$ is the formal solution of
\begin{eqnarray}
G(\omega,\varepsilon)=G^{(+)}_0(\omega)+G^{(+)}_0(\omega)
V(\varepsilon) G(\omega,\varepsilon)\ ,
\label{ocm9}
\end{eqnarray}
the $T$-matrix solution of \eq{ocm5} is derived from
\begin{eqnarray}
G(\omega,\varepsilon)=G^{(+)}_0(\omega)+G^{(+)}_0(\omega)
T(\omega,\varepsilon) G^{(+)}_0(\omega)\ .
\label{ocm10}
\end{eqnarray}
We find
\begin{eqnarray}
T(\omega,\varepsilon)=-{|u\rangle \langle u|
\over \langle u|G^{(+)}_0(\omega)|u \rangle}
+(\omega-H_0)|u\rangle {1 \over \omega-\varepsilon}
\langle u|(\omega-H_0)\ .
\label{ocm11}
\end{eqnarray}
This expression has seemingly a singularity at $\omega=\varepsilon$ in
the second term.
There is, however, no such singularity
for the initial state $|\omega \rangle$ or $|\phi \rangle$:
\begin{eqnarray}
& & T(\omega,\varepsilon) |\omega \rangle =
-{|u\rangle \langle u|\omega \rangle
\over \langle u|G^{(+)}_0(\omega)|u \rangle}\ ,\nonumber \\
& & T(\omega,\varepsilon) |\phi \rangle =-{|u\rangle \langle u|
\phi \rangle \over \langle u|G^{(+)}_0(\omega)|u \rangle}
+(\omega-H_0)|u\rangle \langle u|\phi \rangle\ .
\label{ocm12}
\end{eqnarray}
In particular, the on-shell $T$-matrix given by
\begin{eqnarray}
\langle \omega | T(\omega,\varepsilon) |\omega \rangle
=-{\langle \omega |u\rangle \langle u|\omega \rangle
\over \langle u|G^{(+)}_0(\omega)|u \rangle}\ ,
\label{ocm13}
\end{eqnarray}
has no $\varepsilon$-dependence.
The $T$-matrix elements we need in a nuclear medium is
\begin{eqnarray}
\langle \phi^\prime | T(\omega,\varepsilon) |\phi \rangle
=-{\langle \phi^\prime |u\rangle \langle u|\phi \rangle
\over \langle u|G^{(+)}_0(\omega)|u \rangle}
+(\omega-\varepsilon^\prime)\langle \phi^\prime
|u\rangle \langle u|\phi \rangle\ ,
\label{ocm14}
\end{eqnarray}
where
\begin{equation}
(\varepsilon^\prime - H_0) |\phi^\prime \rangle = 0 \ .
\label{ocm15}
\end{equation}

Summarizing this subsection, we find that, to define
the complete off-shell $T$-matrix,
it is convenient to start with \eq{ocm4},
which is similar to the Schr{\" o}dinger-type equation
for the defect function in the $G$-matrix formalism. \cite{BE63}
The energy $\varepsilon$ in $V(\varepsilon)$ is
the relative energy for the initial two-particle state,
and should not be mixed up with
the energy $\omega$ in the free Green function.
The energy $\omega$ is usually negative and sometimes referred
as the starting energy in $G$-matrix calculations
for ground state properties.
In the former equation \eq{ocm3} the orthogonality
is imposed on $\psi$, while the defect
function $\chi=\psi-\phi$ respects
the orthogonality to the Pauli forbidden state $u$.

\bigskip

\subsection{Noyes-Kowalski equation}

The Lippmann-Schwinger-type equations in the momentum representation
are nicely solved by the techniques
developed by Noyes \cite{NO65} and Kowalski \cite{KO65}.
For details of this method, the original papers should
be referred to.
Here we recapitulate only minimum equations
necessary for the following discussion.
Let us consider the Lippmann-Schwinger equation for the wave-function
\begin{eqnarray}
|\psi^{(+)}\rangle=|\phi\rangle+G^{(+)}_0(E) V |\psi^{(+)} \rangle
\qquad \hbox{with} \qquad (E-H_0)|\phi \rangle=0\ ,
\label{nk1}
\end{eqnarray}
or the $T$-matrix equation 
\begin{eqnarray}
T(E)=V+V G^{(+)}_0(E) T(E)\ .
\label{nk2}
\end{eqnarray}
We assume that the partial-wave decomposition is already made, and write
the free Green function as
\begin{eqnarray}
G^{(+)}_0(E)=\CP {1 \over E-H_0}-i\pi \delta (E-H_0)
=\CP G_0(E) - i\pi |\phi \rangle \langle \phi |\ .
\label{nk3}
\end{eqnarray}
Then, it is convenient to deal with the $K$-matrix equation 
\begin{eqnarray}
R(E)|\phi \rangle = V |\phi \rangle + V \CP G^{(+)}_0(E) R(E)
|\phi \rangle \ ,
\label{nk4}
\end{eqnarray}
instead of $T(E)|\phi \rangle$, which is obtained from
\begin{eqnarray}
R(E)|\phi \rangle = T(E)|\phi \rangle [1-i\pi \langle \phi|T(E)|\phi
\rangle]^{-1} \ .
\label{nk5}
\end{eqnarray}
The essential point of the Noyes-Kowalski formalism is to use
\begin{eqnarray}
W=V-V|\phi \rangle {1 \over \langle \phi|V|\phi \rangle}
\langle \phi |V\ ,
\label{nk6}
\end{eqnarray}
satisfying $\langle \phi|W=0$ and $W|\phi \rangle=0$.
The solution of \eq{nk4} is factorized as
\begin{eqnarray}
R(E)|\phi \rangle = f|\phi \rangle \langle \phi|R(E)|\phi \rangle\ .
\label{nk7}
\end{eqnarray}
Then the square integrable function $|f\rangle=f|\phi \rangle
\langle \phi |V|\phi \rangle$ should satisfy the basic equation
\begin{eqnarray}
|f \rangle = V |\phi \rangle +W G_0(E) |f \rangle\ .
\label{nk8}
\end{eqnarray}
Note that $W G_0(E)$ with $G_0(z)=1/(z-H_0)$ is
the Hilbert-Schmidt kernel, and the solution of \eq{nk8} satisfies
a trivial relationship $\langle \phi|f|\phi \rangle=1$.
The on-shell $T$-matrix is calculated from
\begin{eqnarray}
\langle \phi|R(E)|\phi \rangle = [1-\langle \phi|V \CP G_0(E)f|
\phi \rangle]^{-1}
\langle \phi|V|\phi \rangle\ ,
\label{nk9}
\end{eqnarray}
and
\begin{eqnarray}
\langle \phi|T(E)|\phi \rangle = [1+i\pi \langle \phi|R(E)|\phi \rangle]^{-1}
\langle \phi|R(E)|\phi \rangle .
\label{nk10}
\end{eqnarray}
The half off-shell $T$-matrix is given by
\begin{eqnarray}
T(E)|\phi \rangle = f|\phi \rangle \langle \phi|T(E)|\phi \rangle\ .
\label{nk11}
\end{eqnarray}

In order to obtain the complete off-shell $T$-matrix,
we should generalize \eq{nk8} as
\begin{eqnarray}
f= V \langle \phi|V|\phi \rangle^{-1} +W G_0(E) f\ ,
\label{nk12}
\end{eqnarray}
and its transpose
\begin{eqnarray}
\widetilde{f}=\langle \phi|V|\phi \rangle^{-1} V
+\widetilde{f} G_0(E) W\ .
\label{nk13}
\end{eqnarray}
The full $T$-matrix is given by
\begin{eqnarray}
T(E)=f|\phi \rangle \langle \phi|T(E)|\phi \rangle \langle \phi |\widetilde{f}
+\left\{ \begin{array}{c}
f \langle \phi |V|\phi \rangle - f|\phi \rangle \langle \phi |V \\ [2mm]
\langle \phi |V|\phi \rangle \widetilde{f}
- V|\phi \rangle \langle \phi |\widetilde{f} \\
\end{array} \right\} \ .
\label{nk14}
\end{eqnarray}

Let us discuss the condition for which the solution
of \eq{nk8} exists.
From the Fredholm's alternative theorem,
it is essential to examine if there exist square integrable
functions $\langle \psi |$ for the conjugate homogeneous equation
\begin{eqnarray}
\langle \psi |(1-WG_0(E))=0\ .
\label{nk15}
\end{eqnarray}
If there exist such solutions, the necessary and sufficient
condition for the unique solution of \eq{nk8} is
\begin{eqnarray}
\langle \psi |V| \phi \rangle =0\ ,
\label{nk16}
\end{eqnarray}
for all $\psi$. Suppose there exists only
one $\psi$ for \eq{nk15}, and \eq{nk8} has
a special solution $|f_0 \rangle$. Then the
general solution of \eq{nk8} is given by
\begin{eqnarray}
|f\rangle=|f_0\rangle+C(E-H_0)|\psi \rangle
\label{nk17}
\end{eqnarray}
for any arbitrary constant $C$. If $\psi$ does not satisfy \eq{nk16},
\eq{nk8} is unsolvable. We can also rewrite the integral
equation \eq{nk15} to the form of the differential equation
\begin{eqnarray}
(E-H_0-W)|\psi \rangle=0\ ,
\label{nk18}
\end{eqnarray}
with the proper boundary condition.
The plane wave $|\phi \rangle$ fulfills \eq{nk18}, but
is not a solution of \eq{nk15} since it is not square integrable.
Unfortunately, we can not prove or disprove the existence
of the positive-energy bound state for $W$ in \eq{nk18}, even
for the standard potential $V$.
In the following, we assume that there is no such solution
which satisfied \eq{nk15} for the standard potentials $V$.

When we apply the present formalism
to the simple OCM equation \eq{rgm12} or to the RGM equation \eq{rgm8},
we find an apparent solution $|\psi\rangle=|u\rangle$.
In fact, for $(\varepsilon-H_0)|\phi\rangle=0$,
we find that $W=W(\varepsilon)$ for $V=V(\varepsilon)$ is
given by
\begin{eqnarray}
W(\varepsilon)={(\varepsilon-H_0)|u\rangle \langle
u|(\varepsilon-H_0) \over \varepsilon-\varepsilon_0}
\qquad \hbox{with} \qquad
\varepsilon_0=\langle u|H_0|u\rangle\ .
\label{nk19}
\end{eqnarray}
This $W(\varepsilon)$ satisfies
\begin{eqnarray}
W(\varepsilon)|u\rangle=(\varepsilon-H_0)|u\rangle
\qquad \hbox{and} \qquad \langle u|W(\varepsilon)
=\langle u|(\varepsilon-H_0)\ ,
\label{nk20}
\end{eqnarray}
which is similar to 
\begin{eqnarray}
V(\varepsilon)|u\rangle=(\varepsilon-H_0)|u\rangle
\qquad \hbox{and} \qquad \langle u|V(\varepsilon)
=\langle u|(\varepsilon-H_0)\ .
\label{nk21}
\end{eqnarray}
Thus we can easily see that $|\psi\rangle=|u\rangle$ is a solution
of \eq{nk18}; i.e., $(\varepsilon-H_0-W(\varepsilon))|u\rangle=0$,
and satisfies $\langle u|V(\varepsilon)|\phi \rangle=0$.
When we apply the present formulation to the RGM equation \eq{rgm10},
$W$ for $V=V(\varepsilon)+v$ is no more simple like \eq{nk19},
but the basic relationship \eq{nk20}
for $W(\varepsilon) \rightarrow W$ and $V(\varepsilon)
\rightarrow V$ is still valid
owing to the property $v=\Lambda v \Lambda$.
We assume that there is no other solution $\psi$ for \eq{nk15}.

Let us rederive the $T$-matrix of $\S$2.2 for the simple OCM.
For the on-shell and the half off-shell $T$-matrix, the direct use
of \eq{nk19} in \eq{nk8} allows the solution
\begin{equation}
|f_0 \rangle=(\varepsilon_0-H_0)|u\rangle
\langle u|\phi \rangle\ ,\qquad
C={1 \over \varepsilon-\varepsilon_0}\langle u|f\rangle\ .
\label{nk22}
\end{equation}
Here $f_0$ is a special solution
which satisfies $\langle u|f_0 \rangle=0$.
Using this solution, we can easily derive
\begin{eqnarray}
\langle \phi | R(\varepsilon) |\phi \rangle
=-{\langle \phi |u\rangle \langle u|\phi \rangle
\over \langle u| \CP G_0(\varepsilon)|u \rangle}\ ,\quad
\langle \phi | T(\varepsilon) |\phi \rangle
=-{\langle \phi |u\rangle \langle u|\phi \rangle
\over \langle u| G^{(+)}_0(\varepsilon)|u \rangle}\ .
\label{nk23}
\end{eqnarray}
For the complete off-shell $T$-matrix, we again need
to assume $\omega \neq \varepsilon$, since
otherwise \eq{nk19} applied to \eq{nk12}   
leads to the condition $\langle u|(\varepsilon-H_0)=0$,
which is apparently not satisfied for a general $\varepsilon$.
The decomposition of $V(\varepsilon)$ with respect
to $(\omega-H_0)|\omega \rangle=0$ leads to the result
\begin{eqnarray}
W(\omega,\varepsilon) & = & V(\varepsilon)
-V(\varepsilon)|\omega \rangle
{1 \over \langle \omega |V|\omega \rangle}
\langle \omega |V(\varepsilon)\nonumber \\
& = & {(\omega-H_0)|u\rangle \langle
u|(\omega-H_0) \over 2\omega-\varepsilon-\varepsilon_0}\ .
\label{nk24}
\end{eqnarray}
After some calculations, the solution
of \eq{nk12} with $E$, $|\phi \rangle$,
$V$ and $W$, being replaced with $\omega$, $|\omega \rangle$,
$V(\varepsilon)$ and $W(\omega,\varepsilon)$, respectively,
is found to be
\begin{eqnarray}
f={1 \over \langle \omega |u\rangle \langle u|\omega \rangle} 
\left\{|u\rangle \langle u|
-{(2\omega-\varepsilon-H_0)|u\rangle
\langle u|(\omega-H_0) \over
(2\omega-\varepsilon-\varepsilon_0)(\omega-\varepsilon)}\right\}\ .
\label{nk25}
\end{eqnarray}
This expression leads to some simple relations
\begin{eqnarray}
& & f|\omega \rangle={|u\rangle \over \langle \omega |u\rangle}
\ ,\qquad \langle \omega |f|\omega \rangle=1\ ,\nonumber \\
& & \langle u|f={1 \over \langle \omega |u\rangle
\langle u|\omega \rangle} 
\left\{ \langle u|-{1 \over \omega-\varepsilon}
\langle u|(\omega-H_0) \right\}\ ,\nonumber \\
& & \langle u|f|\omega \rangle={1 \over \langle \omega |u\rangle}\ .
\label{nk26}
\end{eqnarray}
Similarly, the solution of \eq{nk13} is given by
\begin{eqnarray}
\widetilde{f}={1 \over \langle \omega |u\rangle \langle u|\omega \rangle} 
\left\{|u\rangle \langle u|
-{(\omega-H_0)|u\rangle
\langle u|(2\omega-\varepsilon-H_0) \over
(\omega-\varepsilon)(2\omega-\varepsilon-\varepsilon_0)}\right\}\ ,
\label{nk27}
\end{eqnarray}
and
\begin{eqnarray}
& & \langle \omega|\widetilde{f}={\langle u| \over
\langle u|\omega \rangle}
\ ,\qquad \langle \omega |\widetilde{f}
|\omega \rangle=1\ ,\nonumber \\
& & \widetilde{f} |u \rangle={1 \over \langle \omega |u\rangle
\langle u|\omega \rangle} 
\left\{ |u \rangle-{1 \over \omega-\varepsilon}
(\omega-H_0)|u \rangle \right\}\ ,\nonumber \\
& & \langle \omega|\widetilde{f}|u\rangle
={1 \over \langle u|\omega \rangle}\ .
\label{nk28}
\end{eqnarray}
Finally, we use Eqs.\,(\ref{nk14}) with $|\phi \rangle
\rightarrow |\omega \rangle$, (\ref{ocm13}),
and (\ref{nk25}) $\sim$ (\ref{nk28}),
to reconstruct $T(\omega,\varepsilon)$.
The final result is, of course, equal to \eq{ocm11}.

\bigskip

\section{The orthogonality in the $T$-matrix for the simple OCM
equation}

We can use the full expression of the $T$-matrix derived in the
preceding section, to investigate how the idea of
the orthogonality to the Pauli-forbidden state is preserved
in the simple OCM. Let us first consider the $\omega \rightarrow
\varepsilon$ limit in the two expressions in \eq{ocm12}.
These two expressions correspond to the wave-function of the
ordinary scattering problem \eq{ocm3} (with $\varepsilon$ being
a simple parameter) and to the off-shell $T$-matrix
in the $G$-matrix formalism in \eq{ocm14}, respectively.
We find
\begin{eqnarray}
\lim_{\omega \rightarrow \varepsilon} T(\omega,\varepsilon)
|\omega \rangle \neq 
\lim_{\omega \rightarrow \varepsilon} T(\omega,\varepsilon)
|\phi \rangle\ .
\label{or1}
\end{eqnarray}
In the left-hand side of \eq{or1} the wave-function is
orthogonal to the Pauli-forbidden state $|u\rangle$,
while in the right-hand side, the portion corresponding
to the defect function of the $G$-matrix is orthogonal.
When $\omega=\varepsilon$, the solution of \eq{ocm1} has
the ambiguity for any admixture of the $|u\rangle$ component,
which resolves the discrepancy of the two different
half off-shell $T$-matrices in \eq{or1}.
In order to see this, we assume $\omega=\varepsilon$ and
derive the wave-function, leaving the ambiguity of $|u\rangle$.
The solution \eq{nk17} with \eq{nk22} immediately gives
\begin{eqnarray}
f|\phi \rangle & = & \left[\,(\varepsilon_0-H_0)|u\rangle
\langle u|\phi \rangle\
+(\varepsilon-H_0)|u\rangle C\,\right]
{1 \over (\varepsilon_0-\varepsilon)\langle \phi|u \rangle
\langle u|\phi \rangle} \nonumber \\
& = & {|u\rangle \over \langle \phi|u\rangle}
+(\varepsilon-H_0)|u\rangle c\ ,
\label{or2}
\end{eqnarray}
The new parameter $c$ is related to $C$ through
\begin{eqnarray}
c={ 1 \over (\varepsilon_0-\varepsilon)\langle \phi|u \rangle
\langle u|\phi \rangle} \left[ \langle u|\phi \rangle+C \right]\ .
\label{or3}
\end{eqnarray}
On the other hand, the wave-function is derived from
\begin{eqnarray}
|\psi^{(+)}\rangle-|\phi \rangle & = & G^{(+)}_0(\varepsilon)
T(\varepsilon)|\phi \rangle
=G^{(+)}_0(\varepsilon) f|\phi \rangle
\langle \phi|T(\varepsilon)|\phi \rangle\ ,\nonumber \\
|\psi_R \rangle-|\phi \rangle & = & \CP G_0(\varepsilon)
R(\varepsilon)|\phi \rangle
=\CP G_0 (\varepsilon) f|\phi \rangle
\langle \phi|R(\varepsilon)|\phi \rangle\ ,
\label{or4}
\end{eqnarray}
where $|\psi_R \rangle$ is the standing-wave solution
for $\CP G_0(\varepsilon)$.
For simplicity, we use the shorthand notation
\begin{eqnarray}
D(\varepsilon) & = & \langle u| \CP G_0(\varepsilon)|u\rangle
\ ,\nonumber \\
D^{(+)}(\varepsilon) & = & \langle u| G^{(+)}_0(\varepsilon)|
u\rangle = D(\varepsilon) - i\pi \langle u|\phi \rangle
\langle \phi |u \rangle\ ,
\label{or5}
\end{eqnarray}
and express the on-shell $T$-matrix as
\begin{eqnarray}
\langle \phi |R(\varepsilon)| \phi \rangle =
-{\langle \phi|u \rangle \langle u|\phi \rangle
\over D(\varepsilon)}\ ,\qquad
\langle \phi |T(\varepsilon)| \phi \rangle =
-{\langle \phi|u \rangle \langle u|\phi \rangle
\over D^{(+)}(\varepsilon)}\ .\nonumber \\
\label{or6}
\end{eqnarray}
From \eq{or2}, $T(\varepsilon)| \phi \rangle$ etc. are
obtained as
\begin{eqnarray}
T(\varepsilon)| \phi \rangle = f|\phi \rangle
\langle \phi|T(\varepsilon)|\phi \rangle
= -{|u\rangle \langle u|\phi \rangle \over D^{(+)}(\varepsilon)}
+(\varepsilon-H_0)|u\rangle \langle u|\phi \rangle B\ ,
\label{or7}
\end{eqnarray}
where the third parameterization of the $|u\rangle$ component, $B$,
is given by
\begin{eqnarray}
B & = & -c{\langle \phi|u \rangle \over D^{(+)}(\varepsilon)}
\nonumber \\
& = & -{1 \over (\varepsilon_0-\varepsilon)
D^{(+)}(\varepsilon) \langle u|\phi \rangle}
\left[ \langle u|\phi \rangle+C \right]\ .
\label{or8}
\end{eqnarray}
We have various orthogonalities, as shown in Table I,
depending on what values we take for the arbitrary $C$, $c$ or $B$.
These half off-shell $T$-matrices have all equal qualifications
for the solution of \eq{rgm12}.

%
\begin{table}[b]
\caption{
The orthogonality properties
for the simple OCM in \protect\eq{rgm12} or \protect\eq{ocm1}.
}
\label{table1}
\bigskip
\begin{center}
\renewcommand{\arraystretch}{2.0}
\setlength{\tabcolsep}{3mm}
\begin{tabular}{cccc}
\hline
\hline
orthogonality & $C$ & $c$ & $B$ \\
\hline
$\langle u|\psi^{(+)}\rangle=0$ & $-\langle u|\phi \rangle$
& 0 & 0 \\
$\langle u|\psi^{(+)}-\phi \rangle=0$
& $-\left[ 1+(\varepsilon_0-\varepsilon)D^{(+)}(\varepsilon)
\right] \langle u|\phi \rangle$
& $-{D^{(+)}(\varepsilon) \over \langle \phi|u\rangle}$ & 1 \\
$\langle u|\psi_R-\phi \rangle=0$
& $-\left[ 1+(\varepsilon_0-\varepsilon)D(\varepsilon)
\right] \langle u|\phi \rangle$
& $-{D(\varepsilon) \over \langle \phi|u\rangle}$
& ${D(\varepsilon) \over D^{(+)}(\varepsilon)}$ \\
$\langle u|f \rangle=0$
& 0 & ${1 \over (\varepsilon_0-\varepsilon) \langle \phi |u\rangle}$
& $-{1 \over (\varepsilon_0-\varepsilon) D^{(+)}(\varepsilon)}$ \\
[2mm]
\hline
\hline
\end{tabular}
\end{center}
\end{table}

Summarizing this section, we have found
that the Lippmann-Schwinger equation \eq{ocm5} is a very general
equation describing not only the free scattering
but also the correlation for the scattering
in a nuclear medium.
For the simple OCM \eq{ocm1}, the solution of the half
off-shell $T$-matrix depends on how the model is formulated from
the more general equation in which the Pauli-forbidden state
does not exist.  
The second method in \eq{ocm4} seems to be more natural
than \eq{ocm3}, since it has a direct physical meaning
of summing up all the ladder diagrams
in the $G$-matrix formulation.
In this case, the orthogonality to the Pauli-forbidden state
is represented with respect to the correlation function,
instead of the wave-function itself.
 
\bigskip

\section{Modification of the RGM kernel}

The uniqueness of the solution for the Noyes-Kowalski
equation \eq{nk8} is derived from the structure $V(\varepsilon)
=(\varepsilon-H_0)-\Lambda (\varepsilon-H_0)
\Lambda$ for $(\varepsilon - H_0)|\phi \rangle = 0$ and
the relationship $v=\Lambda v \Lambda$.
It does not depend on the explicit
form $v=\Lambda V_{\rm RGM}\Lambda
=\Lambda (V_{\rm D}+G+\varepsilon K)\Lambda$.
We can modify the internal-energy part and reduced masses
of the CCRGM kernel, by using this property.
Namely, we leave $\varepsilon$ in $v$ as it is, and set up
the RGM equation as
\begin{eqnarray}
& & \left( \varepsilon^{\rm exp} - H^{\rm exp}_0 \right) \chi
= \left( V^{\rm exp}(\varepsilon)+v \right) \chi \ ,\nonumber \\
& & V^{\rm exp}(\varepsilon)
=\left( \varepsilon^{\rm exp}-H^{\rm exp}_0 \right)
-\Lambda \left( \varepsilon^{\rm exp}-H^{\rm exp}_0 \right) \Lambda\ ,
\label{mod1}
\end{eqnarray}
where
\begin{equation}
\varepsilon^{\rm exp}=E-E^{\rm exp}_{\rm int}\ ,\qquad
H^{\rm exp}_0=-{\hbar^2 \over 2 \mu^{\rm exp}}\left({\partial \over
\partial \br}\right)^2={\mu \over \mu^{\rm exp}}H_0
\label{mod2}
\end{equation}
are the empirical relative energies and free kinetic-energy
operators, respectively.
We denote the modification of these quantities by
\begin{equation}
\Delta E_{\rm int}=E^{\rm exp}_{\rm int}-E_{\rm int}\ ,\qquad
\Delta H_0=H^{\rm exp}_0-H_0
=\left({\mu \over \mu^{\rm exp}}-1\right) H_0\ ,
\label{mod3}
\end{equation}
and use the notation
\begin{equation}
\Delta G=\Lambda \left( \Delta E_{\rm int}+\Delta H_0 \right)
\Lambda - \left(\Delta E_{\rm int}+\Delta H_0 \right)\ .
\label{mod4}
\end{equation}
The new RGM equation \eq{mod1} is equivalent to
\begin{equation}
\Lambda (\varepsilon^{\rm exp} - H^{\rm exp}_0-V_{\rm RGM})
\Lambda \chi = 0\ ,
\label{mod5}
\end{equation}
if we use $v=\Lambda V_{\rm RGM}\Lambda$. This implies that we
have replaced $\varepsilon$ and $H_0$ in the direct term as
\begin{equation}
\varepsilon \rightarrow \varepsilon^{\rm exp}\ ,
\qquad H_0 \rightarrow H^{\rm exp}_0\ ,
\label{mod6}
\end{equation}
in the allowed model space,
without changing $\varepsilon$ in $v=\Lambda V_{\rm RGM}\Lambda
=\Lambda (V_{\rm D}+G+\varepsilon K)\Lambda$.
In the original form of the RGM equation \eq{rgm8},
it is easy to see that \eq{rgm8} should be modified to 
\begin{equation}
(\varepsilon^{\rm exp}-H^{\rm exp}_0-V_{\rm RGM}-\Delta G) \chi=0\ .
\label{mod7}
\end{equation}
This equation indicates that an extra modification
of adding $\Delta G$ is required,
in addition to the modification \eq{mod6} of the direct term.

As an example, let us consider $\Lambda N$-$\Sigma N (I=1/2)$ CCRGM.
The scattering problem of this system is solved
in Ref.\,\citen{FU00a}, using Lippmann-Schwinger
RGM (LS-RGM) formalism.
Since this system involves a complete
Pauli-forbidden state in the $(11)_s$ $SU_3$ representation
for the $\hbox{}^1S_0$ state, \cite{FU95b}
we need a special care for the treatment
of the Pauli principle. In our previous publications,
the realistic treatment of the reduced mass in the direct term,
using the empirical $\Lambda$ and $\Sigma$ masses,
needed some modification of the exchange kinetic-energy kernel.
The procedure adopted in Ref.\,\citen{FU95b},
multiplying the exchange kinetic-energy kernel $G^{\rm K}$ by
the factor $\sqrt{\mu_\alpha/\mu^{\rm exp}_\alpha}$ from
the bra and ket sides,
as $\sqrt{\mu_\alpha/\mu^{\rm exp}_\alpha}
G^{\rm K}_{\alpha \alpha^\prime}
\sqrt{\mu_{\alpha^\prime}/\mu^{\rm exp}_{\alpha^\prime}}$,
is not actually accurate for the coupled-channel problems.
As the result, the previous calculation has given
a catastrophic resonance behavior
for the $\Lambda N$ $\hbox{}^1S_0$ phase shift
in the low-momentum region around $p_\Lambda \sim 100~\hbox{MeV}/c$,
if we take too many mesh points for the momentum discretization.
We, therefore, use the following two-step
modification for the $\Lambda N$ and $\Sigma N$ reduced masses.
We fist multiply all the channels
by the common $\mu_{\alpha=1}/\mu^{\rm exp}_{\alpha=1}$ factor
for the incident baryon channel, just as done for the
single-channel problem (see Ref.\,\citen{NA95}).
This process is necessary to reduce too strong effect
of the momentum-dependent Darwin term
involved in the Fermi-Breit interaction.
The Pauli principle is exactly preserved
at this stage with respect to the kinetic-energy term.
Next we introduce a small modification for the reduced mass
of the second baryon channel with respect to the direct term.
The modification of the exchange term is carried out
by using $\Delta G$ in \eq{mod4}.
Since the magnitude of $\Delta H_0$ term is at most a few MeV,
the error caused by this approximate treatment
of the exchange term should be more than
one order smaller in comparison with the ''exact'' value
of the exchange kernel (even if it is possible to evaluate).
In practice, we augment the exchange kernel $G(\bq_f,\bq_i)$ with
the $S$-wave Born kernel
\begin{eqnarray}
& & \Delta G^{\rm K}(\bq_f,\bq_i)
=\left(\Delta {\hbar^2 \over 2\mu}\right)
f {9 \over 100}
\left(\begin{array}{cc}
{9 \over 4b^2} & -{10 \over3}q^2_i+{27 \over 4b^2} \\ [2mm]
-{10 \over 3}q^2_f+{27 \over 4b^2}
& -10(q^2_f+q^2_i)+{81 \over 4b^2} \\
\end{array} \right)\ ,\nonumber \\
& & \Delta G^{\rm K}(\bq_f,\bq_i)
=\left(\Delta {\hbar^2 \over 2\mu}\right)
f {9 \over 100}
\left(\begin{array}{cc}
{9 \over 4b^2} & -{10 \over3}q^2_i+{3 \over 4b^2} \\ [2mm]
-{10 \over 3}q^2_f+{3 \over 4b^2} & -{10 \over 9}(q^2_f+q^2_i)
+{1 \over 4b^2} \\
\end{array} \right)\ ,\nonumber \\
\label{mod8}
\end{eqnarray}
for $\Lambda N$-incident and $\Sigma N$-incident scatterings,
respectively.
Here we have defined $\Delta \hbar^2/2\mu
=\hbar^2/2\mu^{\rm exp}_{\Sigma N}
-\hbar^2/2\mu^{\rm exp}_{\Lambda N}$,
$b$ is the harmonic-oscillator width parameter,
and the $S$-wave spatial function $f$ is given by
\begin{equation}
f=f(q_f, q_i)=\left( \sqrt{{8\pi \over 3}}b\right)^3
\exp \left\{-{1 \over 3}b^2(q^2_f+q^2_i)\right\}\ .
\label{mod9}
\end{equation}
The condition $\langle u|V_{\rm RGM}+\Delta G|\phi \rangle=0$ for
the Pauli-forbidden state $|u\rangle$ and $|\phi
\rangle$ satisfying $(\varepsilon^{\rm exp}-H^{\rm exp}_0)
|\phi\rangle=0$ guarantees
the existence of the solution of \eq{mod7}.
If we assume $\Delta G=0$ in \eq{mod7}, we obtain the expression
\begin{equation}
\langle u|V_{\rm RGM}|\phi \rangle = -\langle u|\Delta G|\phi \rangle
=\langle u|\Delta (E_{\rm int}+H_0) |\phi \rangle \ .
\label{mod10}
\end{equation}
This expression is actually 0 if the incident momentum
of $\Lambda$ is below the $\Sigma N$ threshold.
This is because the plane wave
solution $|\phi \rangle$ has the non-zero component only
for the $\Lambda N$ channel
and the $2 \times 2$ matrix $\Delta (E_{\rm int}+H_0)$ has a non-zero
component only for the second diagonal channel.
In this particular case, neglecting $\Delta G$ in the second step
is harmless, even if we modify only the direct term.

The modification of the internal-energy term is not necessary
in the $\Lambda N$-$\Sigma N$ coupled-channel problem in the
isospin basis, since the model parameters are usually fixed
to reproduce the mass difference of the $\Lambda$ and $\Sigma$ in
our previous models, FSS and RGM-H. \cite{FU96a,FU96b,FJ98} 
This is, however, no more valid in the particle-basis
calculation \cite{FU98} to study the charge symmetry breaking,
since a consistent description of the baryon-mass splitting
in terms of the up-down quark mass difference and
the Coulomb energies is not always successful.
It may, therefore, be useful to present a general expression
of the extra Born kernel \eq{mod4} in the following:
\begin{eqnarray}
\Delta G_\aagp(\bq_f,\bq_i) & = & \Delta G^{\rm int}_\aagp(\bq_f,\bq_i)
+\Delta G^{\rm K}_\aagp(\bq_f,\bq_i)\ ,\nonumber \\
\Delta G^{\rm int}_\aagp(\bq_f,\bq_i)
& = & c_\alpha c_{\alpha^\prime} f \left[
-\left(\Delta E^{\rm int}_\alpha
+\Delta E^{\rm int}_{\alpha^\prime}\right)
+\sum_{\alpha^{\prime \prime}}c^2_{\alpha^{\prime \prime}}
\Delta E^{\rm int}_{\alpha^{\prime \prime}}\right]
\ ,\nonumber \\
\Delta G^{\rm K}_\aagp(\bq_f,\bq_i)
& = & c_\alpha c_{\alpha^\prime} f \left[
-\left( q^2_f \Delta {\hbar^2 \over 2\mu_\alpha}
+q^2_i \Delta {\hbar^2 \over 2\mu_{\alpha^\prime}}\right)
+{9 \over 4b^2}
\sum_{\alpha^{\prime \prime}}c^2_{\alpha^{\prime \prime}}
\Delta {\hbar^2 \over 2\mu_{\alpha^{\prime \prime}}}\right]\ ,
\nonumber \\
& &
\label{mod11}
\end{eqnarray}
where $f$ is given in \eq{mod9} and
\begin{equation}
\Delta E^{\rm int}_\alpha=\left(E^{\rm int}_\alpha \right)^{\rm exp}
-\left(E^{\rm int}_\alpha \right)^{\rm cal}\ ,\qquad
\Delta {\hbar^2 \over 2\mu_\alpha}
={\hbar^2 \over 2\mu^{\rm exp}_\alpha}
-{\hbar^2 \over 2\mu^{\rm exp}_{\alpha=1}}\ .
\label{mod12}
\end{equation}
The normalized eigenvector $c_\alpha$ of the
Pauli-forbidden state in the flavor space is obtained from
the eigenvalue equation
\begin{equation}
\sum_{\alpha^\prime}\left(\lambda \delta_\aagp
+\left(X_N\right)_\aagp \right) c_{\alpha^{\prime}}=0
\qquad \hbox{with} \qquad \lambda=1\ ,
\label{mod13}
\end{equation}
by using the spin-flavor-color factors of the exchange normalization
kernel $X_N$. The result of \eq{mod8} is obtained
if one uses $c_1=1/\sqrt{10},~c_2=3/\sqrt{10}$ for
the $\Lambda N$-incident
channel and $c_1=3/\sqrt{10},~c_2=1/\sqrt{10}$ for
the $\Lambda N$-incident channel. This formula can also be used
for the $\Lambda \Lambda$-$\Xi N$-$\Sigma \Sigma$ CCRGM
with the strangeness $S=-2$.

\bigskip

\section{Discussion and summary}

In this paper we have discussed what kind of equation we should
solve for the off-shell $T$-matrix derived
from the $(3q)$-$(3q)$ resonating-group method (RGM) for
the baryon-baryon interaction. This is an important issue since
the present-day quark-model description
for the nucleon-nucleon ($NN$) and
the hyperon-nucleon ($YN$) interactions is  
very accurate, and the realistic calculation of the
hypertriton and baryonic matter etc. using directly
the quark-exchange kernel of these interactions is feasible
to investigate the important off-shell effect and
intricate behavior of the short-range
correlations.\footnote{For the $G$-matrix calculation
of the $NN$ and $YN$ systems, using the model FSS and RGM-H,
see Refs.\,\citen{KO00} and \citen{FU00b}}

\addtocounter{footnote}{1}

The off-shell $T$-matrix in RGM requires a rather involved formulation
when the complete Pauli-forbidden state exists.
Since the relative wave-function can contain an arbitrary
admixture of the redundant components, the half off-shell $T$-matrix
should be defined as a limit of some definite $T$-matrix equation
without this ambiguity. As a possible choice of such equation,
we have proposed a standard $G$-matrix equation which uses
the quark-exchange kernel $V_{\rm RGM}
=V_{\rm D}+G+\varepsilon K$ directly as the driving term
for the infinite sum of all the ladder diagrams.
The relative energy $\varepsilon$ should be taken with respect
to the initial two-particle state.\footnote{In this sense,
the prescription to neglect the $\varepsilon K$ term
in $V_{\rm RGM}$, as in Ref.\,\citen{OR94}, is not correct.}
By using a simplified version
of Saito's orthogonality condition model (OCM), \cite{SA68}
we have argued that this inherent energy-dependence
of the Born kernel is essential to preserve
the major role of the Pauli principle,
which is sometimes represented in the form of
the orthogonality condition to the Pauli-forbidden state. 
In the ordinary formulation of OCM this orthogonality condition
is applied to the relative wave-function, \cite{SA68,SA77}
while in the present $G$-matrix formulation the defect-function part
of the correlation function is orthogonal.
We can also start from the OCM-type RGM equation like \eq{rgm9},
and derive a different type of $G$-matrix equation for
the quark-model interaction. But such a formulation needs
a careful treatment of the Pauli forbidden state, which 
is represented by $V(\varepsilon)$ in \eq{rgm11} in this paper.
The present method is the simplest and the most natural
in the sense that such a term is automatically incorporated
in the structure of the exchange kernel for the normalization
and kinetic-energy terms.

We have also clarified the condition for which the
the integral equation derived by Noyes \cite{NO65} and
Kowalski \cite{KO65} is solvable and has a unique solution
for the most general off-shell $T$-matrix.
For the on-shell and half off-shell $T$-matrices,
the conjugate homogeneous equation has a trivial solution
if the Pauli-forbidden state $|u\rangle$ exists.
The condition $\langle u|V_{\rm RGM}|\phi \rangle=0$ is automatically
satisfied for the plane-wave solution $|\phi \rangle$ in the 
initial channel, as long as
the energy $\varepsilon$ in $V_{\rm RGM}$ is fixed
to the energy of $|\phi \rangle$.
We can use this relationship to test if the
exchange kernel is correctly calculated.
When one needs to modify the exchange kernel,
the modification should be made such a way that
this condition is still satisfied.
An example of this kind of modification is
the correction of the reduced masses
and a small readjustment of the threshold
energies in the coupled-channel RGM (CCRGM). 
In the rigorous framework of the RGM, it is sometimes
difficult to reproduce correct reduced masses and
internal energies starting from a unique Hamiltonian.
In particular, the baryon masses for the inertia mass and
the rest mass are sometimes inconsistent in the framework
of the non-relativistic quark model,
due to the residual $qq$ interaction.
On the other hand, the correct kinematics is mandatory
for the realistic description of the baryon-baryon
interaction. We have shown that, if one uses the empirical
reduced masses and threshold energies, an additional
exchange term $\Delta G(\bq_f,\bq_i)$ should be added
in general to the Born amplitude $V_{\rm RGM}(\bq_f,\bq_i)$.
We have given the explicit expression for $\Delta G(\bq_f,\bq_i)$,
which can be used for $\Lambda N$-$\Sigma N$ and $\Lambda
\Lambda$-$\Xi N$-$\Sigma \Sigma$ CCRGM.


\section*{Acknowledgements}

The authors would like to acknowledge Professor K. Yazaki
for motivating this work.
This work was supported by the Grant-in-Aid for Scientific
Research from the Ministry of Education, Science, Sports and
Culture (No. 12640265).



\end{document}